# Lift force acting on an intruder in dense, granular shear flows

Meng Liu [a], Christoph R. Müller [a*]

[a] Department of Mechanical and Process Engineering, ETH Zurich, Leonhardstrasse 21,8092 Zurich, Switzerland

* Corresponding author email address: muelchri@ethz.ch

**Abstract**

We report a new lift force model for intruders in dense, granular shear flows. Our derivation is based on the thermal buoyancy model of Trujillo & Hermann[L. Trujillo and H. J. Herrmann, Physica A **330**, 519 (2003).], but takes into account both granular temperature and pressure differences in the derivation of the net buoyancy force acting on the intruder. In a second step the model is extended to take into account also density differences between the intruder and the bed particles. The model predicts very well the rising and sinking of intruders, the lift force acting on intruders as determined by discrete element model (DEM) simulations and the neutral-buoyancy limit of intruders in shear flows. Phenomenologically, we observe a cooling upon the introduction of an intruder into the system. This cooling effect increases with intruder size and explains the sinking of large intruders. On the other hand, the introduction of small to mid-sized intruders, i.e. up to 4 times the bed particle size, leads to a reduction in the granular pressure compared to the hydrostatic pressure, which in turn causes the rising of small to mid-sized intruders.

## 1. Introduction

Since the first report by Brown [1], segregation effects in granular systems have received widespread interest among both physicists and engineers owing to their high practical relevance. Particles of different sizes [2], densities [3] and possibly also shapes [4] segregate when excited e.g. via vibration, rotation or gas injection. Segregation is readily encountered in many processing apparatuses such as rotating cylinders [5], hoppers [6] or vibrated beds [2]. In industrial applications segregation is typically an undesirable effect as it counteracts mixing. In addition, segregation in granular media is also commonly observed in nature, e.g. rock avalanches and debris flow [7,8]. Arguably, model systems in which one or multiple intruders are immersed in a granular bed have been studied most frequently and in such systems the so-called "Brazil nut phenomenon" (BNP) has been reported. In the BNP, which is not limited to single intruders, a larger particle rises through a bed of smaller particles under external excitation, typically vibration. Depending on the regularity of the vibrational excitation, the BNP has been explained by percolation [2,9] or convection [10]. The percolation model explains the rising of a larger intruder through a bed of smaller particles via a void filling mechanism. During a vibration cycle first a void is formed underneath the intruder. This void is subsequently filled by the small, surrounding bed particles. At the end of a vibration cycle the intruder falls back to a higher rest position. Through a geometric description of the percolation model, Duran et al. [9] predicted that the rise velocity of the intruder depends on the size ratio of the intruder to the bed particles. For very regular vibrations a convective flow field is established that carries the intruder upwards until it reaches the surface of the bed. The intruder is trapped at the surface as the region close to the walls where downwards motion occurs, is typically smaller than the intruder. In this model, the rise velocity of the intruder is independent of the ratio of the size of the intruder to the bed particles. Although, the two models described above provide some conceptual understanding of segregation in granular media, there is still considerable debate on how to model segregation from a hydrodynamic perspective. Such hydrodynamic models require formulations of the granular counterparts of drag, buoyancy, and in specific cases the Saffman forces.

Using discrete element method (DEM) simulations, Shishodia & Wassgren [11] were among the first to establish an expression for the buoyancy force in a granular system. In their 2D vibro-fluidized bed simulations, periodic boundary conditions were employed to eliminate the contribution of convection. In the absence of a convective pattern the intruder was found to rise to an equilibrium position within the bed (bed position $y$), instead of rising to the top. Making an analogy to the fluid mechanic description of the buoyancy force, i.e. the product of a pressure gradient and the intruder volume they proposed the following expression for the buoyancy force, $F_b$:

$$\mathbf{F}_b = \nabla P(y) V_I = -m_I \mathbf{g}, \qquad (1)$$

where $m_I \mathbf{g}$ is the weight of the intruder, $\nabla P(y)$ is the pressure gradient and $V_I$ is the intruder volume.

It is worth noting that the 2D granular system investigated by Shishodia & Wassgren [11] was in the granular gas regime in which binary particle collisions dominate. When considering practically more relevant dense granular systems in which multi-particle collisions and long-lasting contacts dominate, the buoyancy force predicted through Eq. (1) underestimates the measured buoyancy force acting on the intruder when the intruder size approaches the size of the bed particles [12].

On the other hand, Trujillo & Herrmann [13] developed a granular buoyancy model using the kinetic theory of granular gases. The system considered contained a single intruder in a vibrated bed. The driving force acting on the rising intruder was modelled as a thermally-induced buoyancy force, i.e. a density difference arising from differences in the granular temperature of the system with and without the intruder. In their model, a reference state "0" was defined that is characterized by a granular pressure $P_0(\mathbf{r})$ and temperature $T_0(\mathbf{r})$, where $\mathbf{r}$ is the position of the intruder. The reference state assumes a bed without the intruder. Upon introduction of the intruder, the bed transitions into a new state "1" which is characterized by $P_1(\mathbf{r})$ and $T_1(\mathbf{r})$. The temperature difference between the "perturbed" and reference states is given as $\Delta T(\mathbf{r}) = T_1(\mathbf{r}) - T_0(\mathbf{r})$, yielding the following expression for a granular, thermally-induced buoyancy force [13]:

$$F_b = \alpha \Delta T \rho V_I g, \quad (2)$$

with $\rho$ being the bulk density of the reference state (i.e. $\rho = \rho_p \phi$ with $\rho_p$ being the density of the bed particles and $\phi$ being the solid fraction of the bed), $g$ is the acceleration due to gravity and $\alpha$ is the coefficient of thermal expansion, defined as $\alpha = -\frac{1}{n}\left(\frac{\partial n}{\partial T}\right)_P$, where $n$ is the number density of the bed particles. Assuming $\alpha$ to be constant, Trujillo & Herrmann showed that $\alpha = \frac{1}{T_0}C(\phi)$, where $C(\phi)$ depends on the solid fraction with $C(\phi) \rightarrow 1$ for $\phi \rightarrow 0$. The system considered by Trujillo & Hermann [13] was a vibro-fluidized bed and a uniform system pressure i.e. d$P \sim 0$ was assumed. However, such a simplification would not be valid in dense, shear systems (*vide infra*).

To summarize, the buoyancy models described in Eq. (1) and Eq. (2) have been developed for systems that operate in the granular gas regime. However, when considering more "liquid-like", dense granular systems additional effects have to be considered in the buoyancy model. For example, in an experimental study of a dense, vertically vibrated bed (amplitude $A = 9.76$ mm and frequency $f = 9.7$ Hz) Shinbrot & Muzzio [14] observed that intruders ($d_I = 152$ mm) with a density $< 0.5\rho_p$ sink, whereas heavy intruders ($d_I = 152$ mm) with a density in the range 1.2-1.7$\rho_p$ rise. This unexpected behaviour has been termed reverse Brazil nut phenomenon (RBNP). As the intruder size was fixed in these two experiments, the buoyancy model described in Eq. (1) cannot explain why the heavier intruder rises to the top while the lighter intruder sinks. It has been argued that interstitial air in beds of small particles (< 800 μm) might contribute to the RBNP [14]. To weaken the influence of the interstitial gas in a vibrated bed, Huerta et al. [12] investigated the BNP in a dense bed containing larger particles (i.e. a mixture of glass beads of 3 and 4 mm in diameter). Huerta et al. [12] observed that a light intruder rises faster than a heavier intruder of equal size. Unlike in the setup of Shinbrot & Muzzio [14], the bed of Huerta et al. [12] was vibrated horizontally with neighbouring sidewalls vibrating with the same amplitude but out of phase (phase shift $\pi$), ensuring the cross sectional area of the bed to remain almost constant over a vibration cycle and avoiding in turn the establishment of a convection pattern. Huerta et al. [12] measured the total lift force acting on the intruder by connecting the intruder, placed in the centre of the bed, with a force sensor. The measured total lift force $\mathbf{F}_{\text{lift}} = -(\mathbf{F}_s + \mathbf{F}_g)$, where $\mathbf{F}_g$ is the gravitational force of the intruder and $\mathbf{F}_s$ is the time averaged value obtained from the force sensor, that can be interpreted as the buoyancy force acting on the intruder. The measured lift force was fitted to a generalized Archimedean formulation of the buoyancy force, viz:

$$F_b = |\mathbf{F}_s + \mathbf{F}_g| = \rho_p \phi V_I g, \quad (3)$$

where the "fitting constant" $\phi$ was very close to the average solid fraction of the bed, i.e. the intruder rises as in a fluid with a density that is equal to the bulk density of the granular media ($\phi \rho_p$). However, the experimental data acquired by Huerta et al.[12] showed only good agreement with the buoyancy model given by Eq. (3) for large ($d_I/d_p > 4$) and very light intruders ($\rho_I/\rho_p = 0.0169$). As the intruder size approached the size of the bed particles, the measured buoyancy force exceeded the predictions of the generalized Archimedean principle given in Eq. (3). Since in a convection-free, vibrated bed in which the intruder is fixed in an equilibrium position and only buoyancy and gravity forces are acting on the intruder, the buoyancy force will become smaller than the bed particle weight for $V_I \rightarrow V_p$, i.e. $F_b = \phi \rho_p V_p g < m_p g$ as the solid fraction of the bed $\phi < 1$. Thus, the generalized Archimedean principle expressed in Eq. (3), underestimates the buoyancy force acting on a bed particle. This limitation of the buoyancy model given in Eq. (3) has also been remarked by van der Vaart et al. [15].

Besides the classic BNP in vibrated systems, particles of different sizes segregate also in dense, shear flows. Savage & Lun [16] proposed that in dense shear flows segregation is driven by both kinetic sieving and squeeze expulsion. Overall, there is a higher probability of finding a void into which a small particle can fall compared to a void into which a large particle can fall. This size-dependent, gravity induced segregation mechanism has been termed "random fluctuating sieving" or "kinetic sieving". In addition, a force imbalance on a particle leads to the particle being squeezed out of its layer. This mechanism was termed "squeeze expulsion", but it is neither necessarily size dependent nor does it has a preferred direction. However, there is currently no hydrodynamic model that describes accurately the motion of segregating intruder(s) in dense, granular shear flows. To gain some insight into these systems, Guillard et al. [17] performed 2D, steady-state, shear flow simulation using DEM to quantify the lift force acting on an intruder as a function of the prevailing pressure (and stress) gradient. The size of the intruder was varied from $d_p$ to $10d_p$ while fixing the intruder density to the density of the bed particles. In their simulations, the intruder was kept at a position of half the height of the bed ($h_c/2$) by connecting it to a virtual spring. The virtual spring imposed an additional (spring) force onto the intruder, i.e. $\mathbf{F}_s = -k_s(y_1 - y_0)\mathbf{e}_y$, where $k_s$ is the spring constant, $y_1$ is the vertical position of the intruder at a given time and $y_0 = h_c/2$ is the initial position of the intruder and $\mathbf{e}_y$ is the unit vector in the $y$ direction. The virtual spring ensures the intruder to remain at its equilibrium position while allowing its free movement along the direction of the flow. The buoyancy force acting on the intruder was calculated in analogy to Eq. (1) and expressed as a function of the spatial gradients of the granular pressure and shear, viz:

$$F_b = -\pi \frac{d_I}{4}\left(F(\mu, d_I/d_p)\frac{\partial P}{\partial y} + G(\mu, d_I/d_p)\frac{\partial |\tau|}{\partial y}\right), \quad (4)$$

where $\tau$ is the granular shear stress, $P$ is the granular pressure, and $\mu = |\tau|/P$ is the bulk friction coefficient. Guillard et al. [17] proposed the factors $F$ and $G$ to be exponential functions of $\mu$ and $d_I/d_p$.

In a subsequent study, van der Vaart et al. [15] aimed to elucidate whether the lift force acting on an intruder in a shear flow can be expressed as the sum of a granular buoyancy force and a Saffman-type lift force, i.e.

$$F_{\text{lift}} = F_b + F_{\text{saff}}. \quad (5)$$

where $F_{\text{saff}} = -a_0 b_0 I_\theta \mu^{-0.5}(d_p/d_I - 1)d_I^2 d_p^{-1}\text{sgn}(\dot{\gamma})$ ($a_0$ and $b_0$ are fitting constants, $I_\theta = \dot{\gamma}d_p/(P/\rho_p)^{0.5}$, $\dot{\gamma}$ is the shear rate and $\mu$ is the bulk friction coefficient). Van der Vaart et al. [15] considered a full 3D, dense shear flow along an inclined plane with an inclination angle $\theta$ with respect to the horizontal direction. Van der Vaart et al. [15] modified Eq. (3) to correct for the underestimation of the buoyancy force for $V_I \to V_p$ by replacing the solid fraction of the bed, $\phi$, by $\phi/\phi_I$ where $\phi_I$ is the solid fraction of the intruder, yielding:

$$F_b = \rho_p \frac{\phi}{\phi_I} V_I g. \quad (6)$$

The solid fraction of the intruder is defined as the ratio of its Voronoi volume, $\tilde{V}_I$ to its physical volume $V_I$. The modified buoyancy model Eq. (6) has been also recently assessed and validated in a vibro-fluized system [18]. The derivation of the Saffman lift force is limited to conditions where inertia is not dominating the local flow around the intruder, i.e. when the shear rate-based Reynolds number Re $\ll$ 1. Furthermore, it is currently unclear whether the Saffman lift force model also holds for very large intruders which have been found to sink [19].

In the most recent work, Jing et al. [20] reported a simple buoyancy based model to describe the lift force acting on a single spherical intruder in a dense, granular shear flow. By varying the size and density ratio of the intruder to bed particles, it was found that the total lift force $F_{\text{lift}}$, as determined via a virtual spring, i.e. $\mathbf{F}_{\text{lift}} = -(\mathbf{F}_s + \mathbf{F}_g)$, collapses onto an Archimedean-type model, viz:

$$F_b = f(D)\rho_p \phi V_I g, \quad (7)$$

where $f(D) = (1 - c_1 \exp(-D/a_1))(1 + c_2 \exp(-D/a_2))$ is a fitting function with fitting constants $c_1 = 1.43$, $c_2 = 3.55$, $a_1 = 0.92$, $a_2 = 2.94$ and $D = d_I/d_p$. As for $D \gg 1$ $f(D) \to 1$, the observation that with increasing $D$ the total lift force approaches the generalized Archimedean principle (Eq. (3)) is captured by Eq. (7). This trend of the buoyancy force for D $\gg$ 1 which can be considered as the continuum limit, can be explained as follow. As the surface area of the intruder increased with $D^2$ the number of contacts between the intruder and the surrounding bed particles increases rapidly with increasing $D$, yielding a uniform stress transmission to the intruder for D $\gg$ 1 and hence a similar behaviour as an intruder immersed into a fluid. On the other hand, for $D \to$ 1, the stress distribution on the intruder is highly anisotropic leading to a deviation from Eq. (3). However, the effect of stress anisotropy on the lift force acting on an intruder remains largely unclear. Nonetheless, despite the empirical derivation of $f(D)$, Eq. (7) can be of practical importance allowing to make an *a priori* prediction whether an intruder of a given size and density ratio will sink or rise.

To summarize, recent works have improved significantly our understanding of size-driven segregation of intruders in dense shear flows and its hydrodynamic modelling. However, although several works observe a distortion of the hydrodynamic pressure field, in addition to the granular temperature field, upon the addition of the intruder, the model of Trujillo & Hermann [13] is limited to disturbances in the temperature field, which is most likely insufficient to describe dense granular systems. Furthermore, while several models have been shown to predict well the forces acting on intruders for lower ratios of $d_I/d_p$ (e.g. $d_I/d_p$ < 4 [15]), models that also predict accurately the sinking of very large intruders would be advantageous. Hence, in this work we aim to extend the original work of Trujillo & Hermann [13] to dense shear flow systems by describing local perturbations in both the pressure and temperature field upon the introduction of an intruder. We demonstrate that our model is able to predict well the transition from rising to sinking for very large intruders. Furthermore, we provide some insight on how and why the presence of an intruder modifies the local pressure and temperature field.

## 2. Method

*2.1 Simulation Method*

DEM simulations of the shear flow system considered here were performed using the LIGGGHTS software [21]. In DEM, each particle is modelled as a single entity (Lagrangian approach) and the normal, $F_{n,\alpha\beta}$, and tangential contact forces, $F_{t,\alpha\beta}$, acting between the contacting particles $\alpha$ and $\beta$ are modelled by a Hertzian contact model [22,23]:

$$F_{n,\alpha\beta} = -k_n\sqrt{\frac{1}{4R^*}}\delta_{n,\alpha\beta}^{3/2} + \gamma_{n,\alpha\beta}\sqrt{m^*}\sqrt[4]{\frac{\delta_{n,\alpha\beta}}{4R^*}}u_{n,\alpha\beta}, \quad (8)$$

$$F_{t,\alpha\beta} = -k_t\sqrt{\frac{1}{4R^*}}\delta_{t,\alpha\beta}^{3/2} + \gamma_{t,\alpha\beta}\sqrt{m^*}\sqrt[4]{\frac{\delta_{t,\alpha\beta}}{4R^*}}u_{t,\alpha\beta}, \quad (9)$$

where $k_n$ and $k_t$ are the spring constants in the normal and tangential direction, respectively. Here, $\delta_n$ and $\delta_t$ are the particle overlaps in, respectively, the normal and tangential direction, $\gamma_n$ and $\gamma_t$ are the damping coefficients in, respectively, the normal and tangential direction, $R^*$ is the effective radius given as $R^* = R_\alpha R_\beta/(R_\alpha + R_\beta)$, $m^* = m_\alpha m_\beta/(m_\alpha + m_\beta)$ and $u_{n,\alpha\beta}$ and $u_{t,\alpha\beta}$ are the tangential and normal relative velocities between particles $\alpha$ and $\beta$, respectively. The tangential contact force, $F_{t,\alpha\beta}$, is limited by Coulomb's law, i.e. $F_{t,\alpha\beta} \leq \mu F_{n,\alpha\beta}$ with $\mu$ being the coefficient of friction.

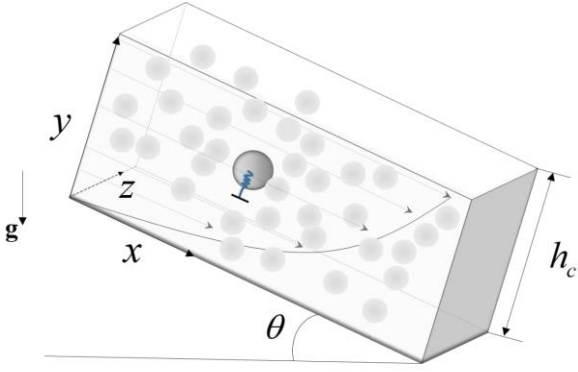

*Figure 1.* Sketch of the granular shear flow system under investigation. An intruder (dark grey) is immersed in the bed at an initial location $y_0 = h_c/2$. The inclination angle is given by $\theta$ and $h_c$ is the distance (in the $y$-direction) from the bottom of the bed to its surface. For simplicity we introduce a local coordinate system. A virtual spring (blue spring) is attached to the intruder. The intruder is able to move freely in the $xz$ plane, while it reaches a (dynamic) equilibrium position $y_1$. The size of the simulation domain is $lx \times lz \times h_c = 20d_p \times 20d_p \times 40d_p$.

Figure 1 illustrates the set-up of the simulation domain. The bed of dimensions $lx \times lz \times h_c = 20d_p \times 20d_p \times 40d_p$ consists of 16'000 bed particles of diameter $d_p$. The bed particles flow along an inclined plane due to gravity. The gravitational vector $|g| = 9.81$ m/s$^2$ can be decomposed into $g_x = g\sin\theta$ and $g_y = g\cos\theta$. Particles with a diameter $d_w = 10$ mm are glued onto the bottom plate to increase the roughness of the bottom wall. In the $x$ and $z$ direction, periodic boundaries were applied to establish a steady-state shear flow. The velocity profiles in the shear flow were varied by adjusting the inclination angle $\theta$. However, as only a narrow range of $\theta$ ensures steady-state conditions, the chute angle $\theta$ was only varied in the range $24° \leq \theta \leq 28°$ in this work. A spherical intruder of diameter $d_I$ was placed inside the bed at a vertical position $y_c = h_c/2 = 20d_p$. The motion of the intruder in the $y$-direction was constrained by a virtual spring (spring constant 80 N/m). The intruder can move freely in the $xz$ plane. The spring force acting on the intruder is determined through its (small) displacement in the $y$ direction, i.e. $\mathbf{F}_s = -k_s(y-y_0)\mathbf{e}_y$. The total lift force acting on the intruder is then given by $\mathbf{F}_{\text{lift}} = -(\mathbf{F}_s + \mathbf{F}_g)$. The parameters used in the DEM simulation are given in Table (1).

Table (1): Parameters used in the DEM simulations

| Parameters | Value |
| --- | --- |
| $k_n$ [N/m] | $6.41 \times 10^4$ |
| $k_t$ [N/m] | $2/7\ k_n$ [24] |
| $d_p$ [m] | 0.005 |
| $d_I$ [m] | $d_p$ up to $\sim 8d_p$ |
| $d_w$ [m] | 0.01 |
| $\rho_p$ [kg/m$^3$] | 2500 |
| $\gamma_n$ [(N/m)$^{1/2}$] | 23.01 [24] |
| $\gamma_t$ [(N/m)$^{1/2}$] | $1/2\ \gamma_n$ |
| $\mu$ | 0.5 |
| $e$ | 0.88 |
| time step [s] | $10^{-5}$ |

*2.2 Coarse Graining*

To obtain the granular pressure, stress and temperature, coarse graining (CG) of the DEM data was employed [25,26]. The granular stress tensor in the coarse graining volume is given by [25,27]:

$$\sigma_{i,j}(\mathbf{r},t) = -\frac{1}{2}\sum_{\alpha,\beta} f_{i,\alpha\beta} r_{j,\alpha\beta} \int_0^1 \Phi(r - r_\alpha + sr_{\alpha\beta}) ds$$
$$- \sum_{\alpha=1}^N m_\alpha \Phi(r - r_\alpha) u'_{i,\alpha} u'_{j,\alpha},$$

(10)

where $\Phi$ is the coarse graining function, $i$, $j$ denotes the Cartesian components, $f_{i,\alpha\beta}$ is the $i$th component of the contact force vector between particles $\alpha$ and $\beta$ (see illustration in Figure 2) and $r_{j,\alpha\beta}$ is the branch vector connecting the centres of gravity of particles $\alpha$ and $\beta$. We use the Heaviside function as the coarse graining function, i.e. $\Phi(\mathbf{R}) = 1/(4/3\pi w^3) H(w - |\mathbf{R}|)$, where $w$ is the radius of the spherical coarse graining volume, $\mathbf{R} = \mathbf{r} - \mathbf{r}_\alpha$ is the vector pointing from a sampling particle inside the coarse graining volume to the center ($\times$) of the coarse graining volume and $u'_{i,\alpha}$ is the velocity fluctuation of particle $\alpha$, viz. $u'_{i,\alpha} = u_{i,\alpha} - \bar{u}_i$ where $u_{i,\alpha}$ is the instantaneous velocity of particle $\alpha$ in the $i$th direction and $\bar{u}_i$ is the average velocity in the $i$th direction of the particles in the coarse graining volume.

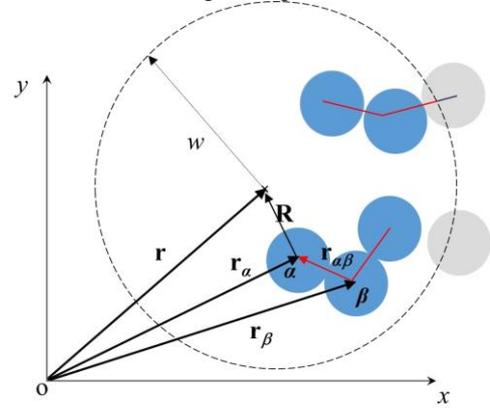

Figure 2. Illustration of the coarse graining method. The coarse graining volume is bounded by a spherical space centred at $\mathbf{r}$ with the coarse graining radius $w$. The branch vector $\mathbf{r}_{\alpha\beta} = \mathbf{r}_\alpha - \mathbf{r}_\beta$ is shown as a red arrow that points from particle $\beta$ to particle $\alpha$.

The average pressure in the coarse graining volume is given by $P = 1/3(\sigma_{ii} + \sigma_{jj} + \sigma_{kk})$. The value of $\sigma_{i,j}$ mainly depends on the coarse graining radius $w$. For example, Figure 3(a) shows the dependence of $\sigma_{yy}$ on the coarse graining radius $w$. For $w/d_p < 1$, $\sigma_{yy}$ increases with increasing $w/d_p$, but reaches an asymptotic value for $w/d_p \geq 1$, in agreement with previous works [26,28]. To avoid an over-smoothing of the local stresses we chose $w = \bar{d} = 1/2(d_I + d_p)$, in agreement with previous works [15,29,30].

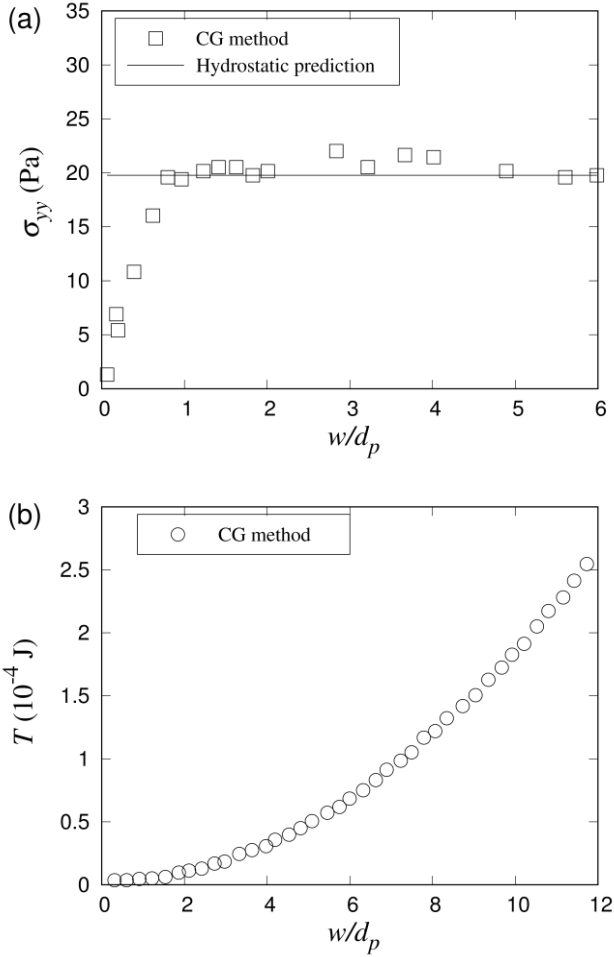

*Figure 3.* Dependence of $\sigma_{yy}$ and the granular temperature on the normalized coarse graining radius $w/d_p$: (a) $\sigma_{yy}$ increases with increasing $w/d_p$ for $w/d_p < 1$. For $w \geqslant d_p$, $\sigma_{yy}$ reaches an asymptotic value that is in excellent agreement with its hydrostatic value, i.e. $Nm_p g/(l_x \cdot l_z)$, where $N$ is the particle number, $m_p$ is the particle mass and $l_x \cdot l_z$ is the cross-sectional area of the system. (b) The granular temperature increases continuously with increasing coarse graining radius $w/d_p$.

Turning now to the granular temperature, viz. [13,31]

$$\frac{3}{2}T = \frac{1}{N}\sum_{\alpha=1}^{N}\frac{1}{2}m_\alpha(u_{i,\alpha}^{'2} + u_{j,\alpha}^{'2} + u_{k,\alpha}^{'2}), \qquad (11)$$

where $i$, $j$ and $k$ denotes the Cartesian components, $u'_{i,\alpha}$, $u'_{j,\alpha}$, $u'_{k,\alpha}$ are, respectively, the velocity fluctuations of particle $\alpha$ with regards to the respective average velocity in the coarse graining volume. The magnitude of the granular temperature depends on the coarse graining volume. For example, in the dense granular shear flow system studied here, the granular temperature increases monotonically with increasing coarse graining volume (Figure 3(b)). Recently, a method has been proposed to eliminate the influence of the coarse graining volume on the granular temperature, however, the approach is only suitable for very specific systems such as monodisperse shear flows [28]. However, as this work concentrates on the effect of a differently sized granular intruder on the granular temperature of the system when compared to the intruder-free reference case, this method is not applicable to the system studied here. Generally, there is very little consensus on the "correct" coarse graining radius for the granular temperature and Glasser & Goldhirsch [32] emphasize to clearly state the coarse graining radius that has been chosen to calculate the granular temperature for a given problem. In the work of Trujillo & Hermann [13] a coarse graining radius of $w = L/3$ ($L$ is the width of the vibrating bed) was chosen for the granular temperature calculation to achieve a good agreement between their thermal buoyancy model and the experimental measurements. Here, we use a coarse graining radius of $w = r_I + d_p$ for the granular temperature, i.e. a value that is very close to the coarse graining radius used for the granular pressure.

## Model Description

Hermann [33] proposed a thermodynamic formulation for granular media that was subsequently adopted to investigate the BNP problem [13]. First, they defined a reference state of a vibro-fluidized bed that is described by a given granular pressure $P_0(\mathbf{r})$ and granular temperature $T_0(\mathbf{r})$. When introducing an intruder into the system, the state at $\mathbf{r}$ changes and is referred to as a perturbed state "1" described by the granular pressure $P_1(\mathbf{r})$ and granular temperature $T_1(\mathbf{r})$. Trujillo & Hermann [13] argued that the perturbed granular system tends to re-establish its reference state, leading to a displacement of the intruder from its initial position $\mathbf{r}$. Neglecting changes in the granular pressure due to the presence of the intruder, Trujillo & Hermann [13] proposed a thermal-driven buoyancy force model (Eq. (2)) to describe the motion of an intruder in a vibro-fluidized bed.

In the following we derive a granular buoyancy model that takes into account also intruder-induced variations in the pressure field to allow the description of the motion of an intruder in a dense, granular shear flow system.

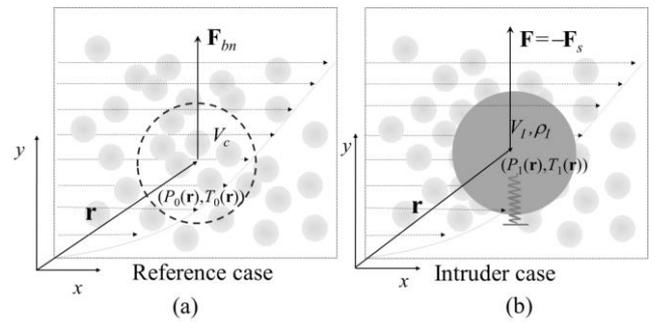

*Figure 4.* Illustration of the reference case and the perturbed state, i.e. the shear flow system with the presence of an intruder. (a) Reference case: Shear flow system without the intruder. The intruder is replaced by an imaginary control volume of size $V_c = V_I$, that is filled with bed particles. The boundary of the imaginary control volume is given by the dashed circle. Here, $P_0(\mathbf{r})$ and $T_0(\mathbf{r})$ denote the reference state at $\mathbf{r}$. (b) Perturbed state: Intruder of volume $V_I$ and density $\rho_I$ is placed at $\mathbf{r}$ into the shear flow system. $P_1(\mathbf{r})$ and $T_1(\mathbf{r})$ are the granular pressure and temperature at $\mathbf{r}$ in the perturbed state. $\mathbf{F}_{bn}$ is the net buoyancy force arising from the temperature and pressure difference between the reference and disturbed states. $\mathbf{F}$ is the force which is balanced by the (virtual) spring force, $\mathbf{F} = -\mathbf{F}_s$ keeping the intruder in the DEM simulations in an equilibrium position.

To quantify the impact of an intruder on the local granular temperature and pressure in a shear flow, we consider the two systems (i.e. the reference and perturbed states) illustrated in Figure 4. Figure 4 (a) sketches the shear flow system without an intruder, i.e. the "reference case", while Figure 4(b) contains an intruder at position $\mathbf{r}$ and is referred to as the "intruder case" (perturbed state). Figure 4(a) is identical to Figure 4(b) except that the intruder is replaced by bed particles. In the reference case, the granular temperature and pressure at $\mathbf{r}$ (i.e. the center of the imaginary control volume) are referred to as $T_0(\mathbf{r})$ and $P_0(\mathbf{r})$, respectively. Similarly, in the intruder case the granular temperature and pressure at $\mathbf{r}$, i.e. the position where the intruder is located, are denoted as $T_1(\mathbf{r})$ and $P_1(\mathbf{r})$, respectively. $P_0(\mathbf{r})$, $T_0(\mathbf{r})$, $P_1(\mathbf{r})$ and $T_1(\mathbf{r})$ are determined through coarse graining. The change in the system due to the presence of an intruder (compared to the reference state) is described by $\Delta T(\mathbf{r}) = T_1(\mathbf{r}) - T_0(\mathbf{r})$ and $\Delta P(\mathbf{r}) = P_1(\mathbf{r}) - P_0(\mathbf{r})$. We follow now the argumentation of Trujillo & Hermann [13] that the thermodynamic driving force for an intruder to sink or rise (i.e. the lift force acting on the intruder) is related to $\Delta T(\mathbf{r})$ and $\Delta P(\mathbf{r})$. As in our shear flow system periodic boundary conditions are applied in the $x$ and $z$ directions, $P(\mathbf{r})$ and $T(\mathbf{r})$ are independent of $x$ and $z$ for a given $y$, allowing us to simplify our notation to $P(y)$ and $T(y)$.

For an inelastic, hard-sphere system (Figure. 4(a)) the granular pressure can be expressed as [34,35]:

$$P = nT\left[1 + \frac{\pi}{3}(1+e)nd_p^3 C(\phi)\right], \quad (12)$$

where $n$ is the particle number density, $T$ is the granular temperature, $e$ is the coefficient of restitution, $d_p$ is the diameter of the particles, $\phi$ is the solid fraction, $C(\phi)$ is the pair correlation function at a contact, i.e., the probability density to find another particle at a distance $d_p$ from a particle center. For a dilute or moderately dense system (i.e. $nd_p^3 \sim 1$), the expression of Carnahan-Starling holds, i.e. $C(\phi) = (2-\phi)/2(1-\phi)^3$ [27] where Carnahan-Starling [27] assumes that binary collisions dominate. This assumption might become inaccurate for denser systems, but we will demonstrate further below that the exact formulation of $C(\phi)$ does not affect appreciably our model predictions. Eq. (12) can be re-formulated as:

$$n = n(P,T). \quad (13)$$

Taking the total differential yields

$$dn = \left(\frac{\partial n}{\partial T}\right)_P dT + \left(\frac{\partial n}{\partial p}\right)_T dP = -\alpha n dT + k_p n dP, \quad (14)$$

where $\alpha$ is the thermal expansion coefficient and $k_p$ is the compressibility coefficient, given by:

$$\alpha = -\frac{1}{n}\left(\frac{\partial n}{\partial T}\right)_P, \quad (15)$$

and

$$k_p = \frac{1}{n}\left(\frac{\partial n}{\partial P}\right)_T. \quad (16)$$

Placing an intruder in the reference shear flow system changes the granular pressure and temperature at position $\mathbf{r}$ from $(T_0, P_0)$ to $(T_1, P_1)$. Assuming $\alpha$ and $k_p$ to be constant and integrating Eq. (14) we obtain,

$$n_1 = n_0 e^{(-\alpha\Delta T + k_p\Delta P)}, \quad (17)$$

where $n_0 = N_0/V_c$ is the number density in the reference case ($N_0$ is the number of particles in the imaginary control volume $V_c$). The density of the control volume can be expressed as $\rho_0 = (N_0 m_p)/V_c = n_0 m_p$, where $m_p$ is the mass of a bed particle. Using the equivalent expression for $\rho_1$, i.e. $\rho_1 = n_1 m_p$, we can rewrite Eq. (17) yielding:

$$\rho_1 = \rho_0 e^{(-\alpha\Delta T + k_p\Delta P)}. \quad (18)$$

Hence, transitioning from state "0" to "1" leads not only to a change in the granular temperature and pressure ($\Delta T$ and $\Delta P$) at position $\mathbf{r}$, but also to a change in density i.e. $\Delta\rho = (\rho_0 - \rho_1)$. We now make the further assumption that the bulk density outside the control volume region (Figure 4(a)) is unaffected by the pressure/temperature perturbation (i.e. it is $\rho_0$). Following Archimedean's principle, we define the net buoyancy force acting on the imaginary control volume in Figure 4(a) as:

$$F_{bn} = (\rho_0 - \rho_1)V_I g_y. \quad (19)$$

As our derivation starts from Eq. (12), which only holds for monodisperse particle systems, the net buoyancy force given in Eq. (19) that is acting on the imaginary control volume in Figure 4(a) is not expected to be the exact equivalent of the lift force acting on the intruder, but it is expected that there exists a strong correlation between $F_{bn}$ and $F_s$ which will be demonstrated in the following.

Substituting Eq. (18) into Eq. (19), and replacing $\rho_0$ with $\rho_p\phi$ we obtain the following expression for the net buoyancy force:

$$F_{bn} = (1 - e^{-\alpha\Delta T + k_p\Delta P})\rho_p\phi V_I g_y. \quad (20)$$

Further, in a steady-state, dense, granular shear flow, the granular pressure at a given height $y$ is given by [36,37]

$$P(y) = (\partial P/\partial y)(y - h_c) = \rho_p\phi g_y(h_c - y), \quad (21)$$

where $h_c$ is the height of the flowing layer. The linear relationship between $P$ and $y$ is confirmed in Figure 5(a).

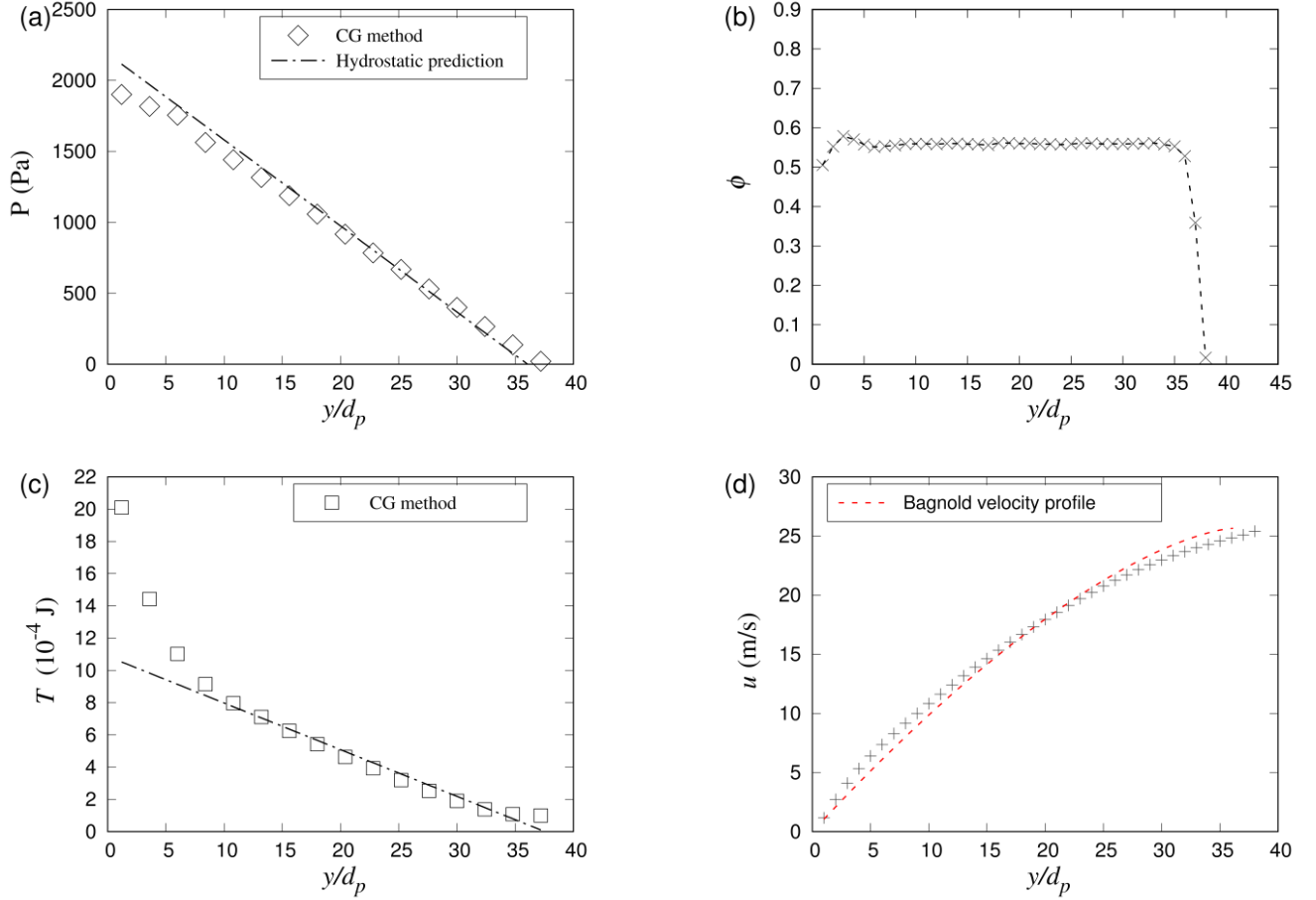

*Figure 5.* (a) Granular pressure along the y direction: (◇) granular pressure as determined by coarse graining (CG), (—·—) hydrostatic pressure i.e. $P_y = \rho_p \phi g_y(h_c - y)$ (Eq. (21)) with $\phi = 0.55$; (b) Solid fraction along the y direction. The solid fraction is nearly constant across the bed ($\phi = 0.55$) except in the regions close to the surface and bottom (distances of approximately $5d_p$); (c) (□) granular temperature (Eq. (11)) as determined by CG, (—·—) linear fit of the CG data, suggesting a linearity between $T$ and $y$ (Eq. (23)) in the core region of the dense, granular, shear flow; (d) Velocity in the x direction as a function of $y/d_p$: (+) DEM data and (---) Bagnold velocity profile [38], i.e. $u(y) = \frac{2}{3} I_\theta \sqrt{P/(\rho_p d_p)} h_c (1-(1-(y/h_c))^{1.5})$, where $I_\theta = \dot{\gamma} d_p/(P/\rho_p)^{0.5}$ (e.g. in this work $I_\theta = 0.17$ for $\theta = 25°$), $P$ is the pressure at the bottom of the system and $h_c$ is the height of the shear flow system. All data are extracted from a shear flow system with an inclination angle $\theta = 25°$.

Substituting now $\phi = nV_p$ into Eq. (21) and combining it with Eq. (16), we obtain the compressibility coefficient:

$$k_p = \frac{1}{n} \frac{1}{m_p g_y(h_c-y)} = \frac{1}{P}. \quad (22)$$

In a steady-state, dense, shear flow system, the solid fraction (and hence also the number density) is constant in the core region [29,36,39] as demonstrated in Figure 5(b). Following Eq. (12), i.e. $P = nTf(e,\phi)$, the granular temperature is also expected to vary linearly with $y$ in regions where $n$ and $\phi$ are constant. This behavior is confirmed in Figure 5(c). It is worth noting that both the granular pressure and temperature, Figures 5(a) and (c), as determined by coarse graining deviate from their linear dependencies with $y$ close to the bottom wall (i.e. at $y < 5d_p$). This wall-induced deviation is in agreement with previous reports, e.g. [40]. However, in our work the intruder is placed well away from the bottom plate and hence, the linear relationship of both the granular pressure and temperature with $y$ is assumed to hold. Thus, combining Eq. (12) and (21) yields:

$$T = \frac{P}{nf(e,\phi)} = \frac{\rho_p \phi g_y(h_c-y)}{nf(e,\phi)}. \quad (23)$$

As $n$, $e$, $\phi$ are constant along $y$ (and away from the boundaries) we obtain:

$$\frac{\partial T}{\partial y} = \frac{1}{nf(e,\phi)} \frac{\partial P}{\partial y} = -\frac{\rho_p \phi g_y}{nf(e,\phi)}. \quad (24)$$

Rearranging Eq. (24) and combining it with Eq. (23) yields:

$$\frac{\partial P}{\partial T} = \frac{\partial P}{\partial y} \bigg/ \frac{\partial T}{\partial y} = n f(e,\phi). \quad (25)$$

In addition from Trujillo & Hermann [13] we have:

$$\frac{\alpha}{k_p} = \left(\frac{\partial P}{\partial T}\right)_n. \quad (26)$$

Combining Eqs. (22), (23) and (26) gives:

$$\alpha = k_p \left(\frac{\partial P}{\partial T}\right)_n = \frac{1}{P} n f(e,\phi) = \frac{1}{T}. \quad (27)$$

Here, following Trujillo & Hermann [13], we have assumed that $k_p$ and $\alpha$ are constant when integrating Eq. (14). Hence, setting $k_p = 1/P_0$ and $\alpha = 1/T_0$, we obtain the following expression for the net buoyancy force acting on the control volume $V_c$ in the reference case (Figure 4(a)):

$$F_{bn} = \left(1 - e^{-\frac{\Delta T}{T_0} + \frac{\Delta P}{P_0}}\right) \rho_p \phi\, g_y V_I. \quad (28)$$

# 3. Results and Discussion

## A. The Lift Force Model

In the following, we first establish the correlation between the net lift force $\mathbf{F} = -\mathbf{F}_s$ acting on the intruder (i.e. $F_s$ is the spring force that prevents the intruder in the DEM simulations from migrating to the top of the shear flow system) and the derived net buoyancy force acting on the imaginary control volume in Figure 4(a), i.e. Eq. (28). Figure 6 plots the normalized spring force, $F_s/m_I g_y$, over the normalized net buoyancy force, $F_{bn}/m_I g_y$, for a series of DEM simulations in which the inclination angle and the size ratio $d_I/d_p$ was varied. From Fig. 6 we obtain a linear correlation between $F_s$ and $F_{bn}$, i.e.

$$F_s = \frac{1}{a} F_{bn}, \quad (29)$$

with $a = 0.55 \pm 0.035$. The bulk solid fraction of our shear flow system is $\phi = 0.54$, hence $\phi/a \cong 1$. Substituting Eq. (28) into Eq. (29) and using $\mathbf{F}_{\text{lift}} = -(\mathbf{F}_s + \mathbf{F}_g) = -(\mathbf{F}_{bn}/a + \mathbf{F}_g)$ the spring force acting on the intruder is given by:

$$F_s = \left(1 - e^{-\frac{\Delta T}{T_0} + \frac{\Delta P}{P_0}}\right) \rho_p g_y V_I. \quad (30)$$

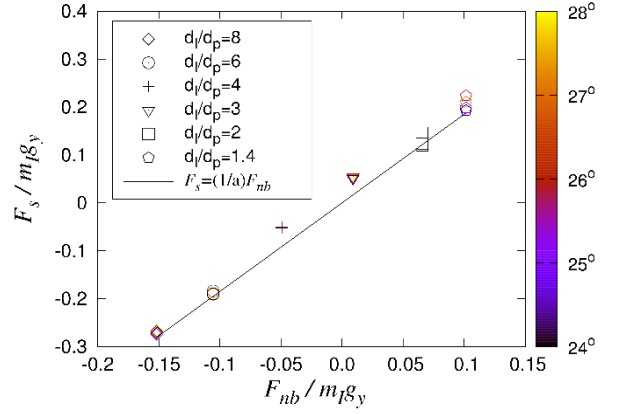

*Figure 6.* Linear relationship between the net buoyancy force, $F_{bn}$, as given by Eq. (28) and the virtual spring force $F_s$. Each data point represents a DEM simulation of a different $d_I/d_p$ ratio (denoted by the marker symbol) and inclination angle ($\theta = 24°, 25°, 26°, 27°, 28°$ denoted by the colour scheme). The density ratio $\rho_I/\rho_p = 1$ was used in the simulations.

Figure 6 shows that the inclination angle has only a very minor influence on the magnitude of the spring force, in agreement with previous reports [15,17,20]. On the other hand, the size ratio $d_I/d_p$ affects the buoyancy force and hence also the spring force, appreciably, i.e. with increasing $d_I/d_p$ the normalized buoyancy force (and the spring force) decrease. To compare our buoyancy model to previously proposed models, we calculate a lift force. The lift force is the sum of all forces that act in the opposite direction of gravity on the intruder (e.g. Saffman and buoyancy forces). For our buoyancy model, the lift force in the specific shear system studied here, is given by:

$$\begin{aligned} F_{\text{lift}} &= \left|-(\mathbf{F}_s + \mathbf{F}_g)\right| \\ &= \left(1 - e^{-\frac{\Delta T}{T_0} + \frac{\Delta P}{P_0}}\right) \rho_p g_y V_I + \rho_p g_y V_I \\ &= \left(2 - e^{-\frac{\Delta T}{T_0} + \frac{\Delta P}{P_0}}\right) \rho_p g_y V_I. \end{aligned} \quad (31)$$

Normalizing the lift force by the gravitational force of the intruder ($\rho_p = \rho_I$) yields:

$$\frac{F_{\text{lift}}}{\rho_I g_y V_I} = 2 - e^{-\frac{\Delta T}{T_0} + \frac{\Delta P}{P_0}}. \quad (32)$$

In the following we performed a series of DEM simulations with varying ratios of $d_I/d_p$ and determined the total lift force acting on the intruder through $\mathbf{F}_{\text{lift}} = -(\mathbf{F}_s + \mathbf{F}_g)$. The differences in the granular pressure and temperature between the reference and the intruder cases, as required for our buoyancy model Eq. (32), were obtained from the Lagrangian DEM data through coarse graining.

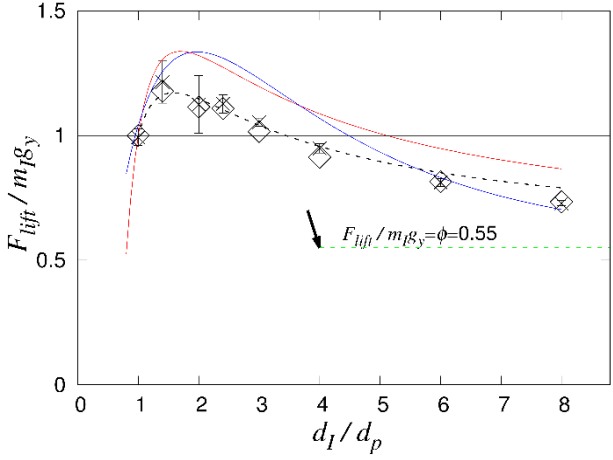

*Figure 7.* Normalized lift force as a function of the size ratio $d_I/d_p$ with $\rho_I/\rho_p = 1$: ($\times$) Lift force acting on the intruder as determined directly from the DEM simulations via a virtual spring. The error bars give the standard deviation obtained from five simulations (each 50 s simulation time in the steady state). ($\diamond$) Lift force as determined by the proposed buoyancy model i.e. Eq. (32). (—) Predictions of the buoyancy model of Jing et al. [19], i.e. Eq. (7) and (—) Saffman force-based lift force model $F_{\text{lift}} = F_{\text{saff}} + F_b = -a_0 b_0 I_\theta \mu^{-0.5}(d_p/d_I - 1)d_I^2 d_p^{-1} \text{s}(\dot{\gamma}) + (\phi/\phi_I)\rho_p g_y V_I$ using $a_0 = 0.24$ and $b_0 = 130.0$ as the fitting parameter as given by van der Vaart et al. [15]. The black dashed line is a prediction of Eq. (5) that uses $a_0 = 0.24$ and $b_0 = 93$ as obtained from our simulation data. The solid, horizontal black line is a guiding reference for $F_{\text{lift}}/(\rho_I V_I g_y) = 1$, i.e. below this reference line an intruder sinks, while values above the reference line indicate a rising intruder. (----) Normalized lift force in the continuum limit ($d_I/d_p \gg 1$), i.e. $F_{\text{lift}}/(\rho_I V_I g_y) \rightarrow \phi \sim 0.55$. In the DEM simulations the inclination angle was varied in the range $\theta = [24°, 28°]$, with $\rho_I/\rho_p = 1$.

Figure 7 plots the lift force determined by the virtual spring, the newly proposed buoyancy model (Eq. (32)), the Saffman-based lift force model (Eq. (5)) and the Archimedean-type buoyancy model given by Eq. (7). Concerning the general trend of the lift force, starting from $d_I/d_p = 1$ where $F_{\text{lift}}/(\rho_I V_I g_y) = 1$, the normalized lift force reaches a maximum at $d_I/d_p \sim 1.5$. The existence of a maximum in the (normalized) lift force with $d_I/d_p$ has been observed previously. For example, Guillard et al. [17] observed a maximum in the lift force at $d_I/d_p \sim 2$ in a 2D plane driven shear flow. Similarly van der Vaart et al. [15] and Jing et al. [20] observed a maximum in the lift force at $d_I/d_p = 1.5$ in a 3D shear chute flow. The reason for the maximum in the lift force is currently unclear, but further below we provide a tentative explanation. For intruder sizes $d_I/d_p > 4$, $F_{\text{lift}}/(\rho_I V_I g_y) < 1$, i.e. the intruder sinks. In several experimental works the sinking of large intruders ($d_I/d_p > 5$ for $\rho_I/\rho_p = 1$) has been observed e.g. in rotating cylinders [19,41].

Our lift force model, Eq. (32), predicts the lift force determined by a virtual spring very accurately, while the Archimedean-type buoyancy model (Eq. (7)) of Jing et al. [20] captures the overall trend well but tends to over-predict the DEM data for $d_I/d_p < 6$. Also the van der Vaart et al. [15] model, Eq. (5), captures very well the overall shape of the lift-force dependency on $d_I/d_p$, but also tends to over-predict the DEM data. When adjusting the fitting parameters of the van der Vaart et al. [15] model to $a_0 = 0.24$ and $b_0 = 93$, a very good agreement with our DEM data is obtained, as illustrated by the dashed line in Figure. 7.

To assess also the dependence of the lift force acting on the intruder on the vertical position of the intruder, the position of the intruder was varied from $y_{\min} = 7d_p$ to $y_{\max} = 34d_p$. According to the velocity profile in the shear flow system under investigation, Figure 5(d), the shear rates vary between 186.0 s$^{-1}$ (at $y = 7d_p$) and 55.5 s$^{-1}$ (at $y = 34d_p$). Figure 8 shows that the lift force acting on the intruder is not sensitive to the vertical position at which the intruder is placed, suggesting that the total lift force acting on the intruder is not sensitive to the shear rate, in agreement with previous studies [15,17,20]

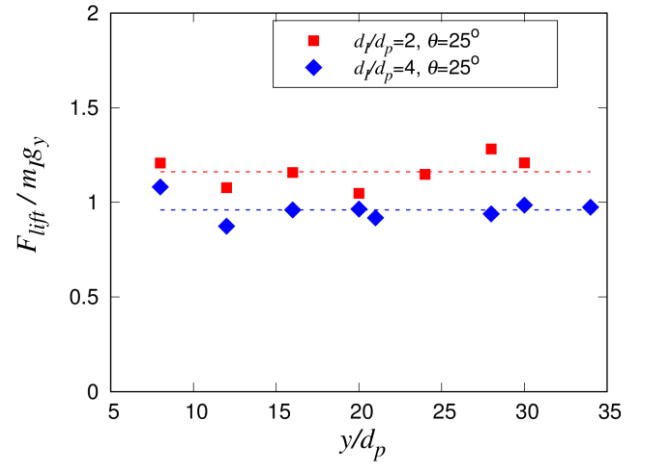

*Figure 8.* Total lift force acting on the intruder determined directly from the virtual spring as a function of intruder position in the $y$ direction (varied from $y_{\min} = 7d_p$ to $y_{\max} = 34d_p$). From the velocity profile given in Figure 5(d) the shear rates vary from 186.0 s$^{-1}$ (at $y = 7d_p$) to 55.5 s$^{-1}$ (at $y = 34d_p$).

### B. Cooling Effect of the Intruder

To elucidate the contributions of the variations of the pressure and temperature field upon addition of an intruder on the lift force, Figure 9 plots $\Delta T/T_0$ and $\Delta P/P_0$ as a function of $d_I/d_p$. We observe that $\Delta T/T_0$ is negative, i.e. the introduction of an intruder leads to a local cooling of the granular temperature. As the ratio $d_I/d_p$ increases the cooling effect becomes stronger. In the following we provide a tentative explanation for the cooling effect of the intruder in dense, shear flows. Concerning the granular system at hand, the major contribution to the granular temperature arises from velocity fluctuations along the shear direction $x$, i.e. $T_x \sim 10^{-5}$ J, $T_y, T_z \sim 10^{-6}$ J for $d_I/d_p = 8$. Hence in the following we focus on the velocity along the $x$ direction. In the reference case, Figure 10(a), assuming a constant shear rate $\dot{\gamma}_0$ in a coarse graining (CG) volume, the granular temperature can be written as, $T_0 \approx \sum_{y=y_0-w}^{y_0+w}$

$(1/(3N))m_p[(\dot{\gamma}_0(y-y_0)+u_0)-u_0]^2$, where $u_0$ is the average particle velocity in the CG volume (dashed circle in Figure 10(a)), N is the number of particles in the coarse graining volume, $m_p$ is the mass of the bed particles and $y_0$ is vertical position of the centre of the CG volume. As $\sum_{y=y_0-w}^{y_0+w}(1/N)[\dot{\gamma}_0^2(y-y_0)^2] = \dot{\gamma}_0^2\langle(y-y_0)^2\rangle$, $T_0$ can be re-written as, $T_0 \approx (1/3)\dot{\gamma}_0^2 m_p \langle(y-y_0)^2\rangle$, where $\langle\rangle$ denotes the average operation in the CG volume (located at $y_0$ with radius $w$). From Figure 3(b) we observe that the granular temperature indeed grows quadratically with $w$, i.e. $T_0 \sim \dot{\gamma}_0^2 w^2$. Hence, two main factors affect the magnitude of the granular temperature: (i) the size of the CG volume i.e. $w$ (Figure 3(b)) and (ii) the magnitude of the shear rate.

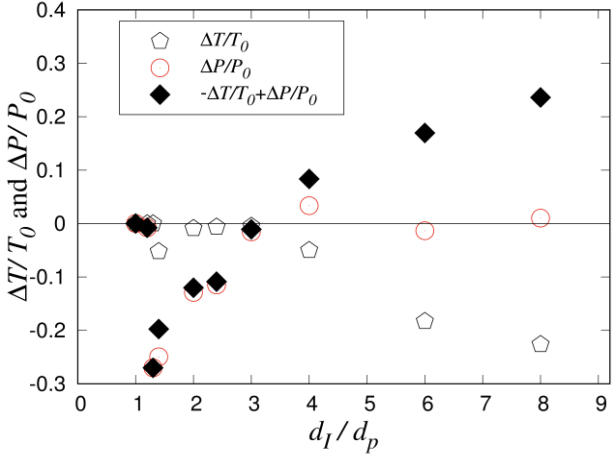

*Figure 9.* Variation of $\Delta P/P_0$ ($\Delta P = P_1-P_0$) and $\Delta T/T_0$ ($\Delta T = T_1-T_0$) as a function of $d_I/d_p$. Both the granular pressure and temperature are determined through coarse graining of the DEM data (simulation performed with an inclination angle $\theta = 25°$ and $\rho_I/\rho_p = 1$). $P_0$ is the hydrostatic pressure that can be determined by coarse graining or analytically ($P_0 = \rho_p g_y(h_c-y_0)$) with $y_0 = 0.1$ m).

Introducing an intruder into the shear system (Figure 10(b)) affects the average velocity in the coarse graining volume (new average velocity $u_1$). Similarly to the reference case (assuming again a constant shear rate in the coarse graining volume) we can write the granular temperature in the intruder case as $T_1 = (1/3)\dot{\gamma}_1^2 m_p \langle(y-y_0)^2\rangle$. Figure 10 plots the velocity of particles (in the shear direction $x$) for the reference and the intruder case, including fits for the shear rate assuming a constant shear rate in the coarse graining volume. For the reference and the intruder cases, shear rates of $\dot{\gamma}_0 = 59.5$ s$^{-1}$ and $\dot{\gamma}_1 = 37.1$ s$^{-1}$ (at $y = 0.1$ m) are obtained, respectively. The shear rate in the intruder case is significantly lower than in the reference case, resulting in a lower granular temperature (cooling) in the intruder case. Overall, it appears that the presence of an intruder leads to a reduced particle velocity at its top and a higher particle velocity at its bottom when compared to the reference case.

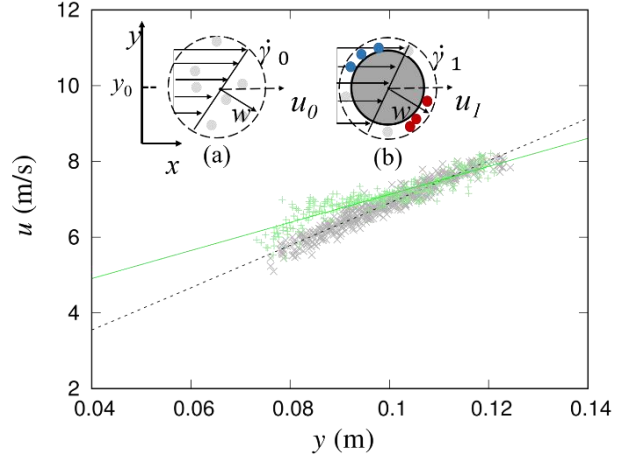

*Figure 10.* Velocity (in the shear direction $x$) of individual particles at $t = 399$ s that are located in the coarse graining volume centered at $y_0 = 0.1$ m. The black dashed circle denotes the CG volume of radius $w = r_I + d_p$ where $r_I = 8d_p$. ($\times$) Particle velocities in the reference case (insert Figure (a)) with (----) being the fitting assuming a constant shear rate in the CG volume, i.e. $\dot{\gamma}_0 = 59.5$ s$^{-1}$ (+) Particle velocities in the intruder case (insert Figure (b)) with (—) being the constant shear rate fitting, i.e. $\dot{\gamma}_1 = 37.1$ s$^{-1}$ in the CG volume. System parameters are $d_I/d_p = 8$, $\rho_I/\rho_p = 1$ and $\theta = 28°$. Inset (a): Coarse graining (CG) volume of the reference case. The CG volume is a spherical space of radius $w$. Here, $u_0$ denotes the average velocity of the particles that are in the CG volume and $\dot{\gamma}_0$ is the shear rate in the CG volume. Inset (b): CG volume of the intruder case, the dark grey area denotes the intruder, $\dot{\gamma}_1$ is the shear rate in the CG volume and $u_1$ is the average velocity of the particles in the CG volume. The blue particles at the top of the intruder indicate a reduced velocity compared to the reference case while the red particles at the bottom of the intruder denote faster particles compared to the reference case.

As a consequence, we observe that the intruder flattens the velocity profile in the CG volume which results in an overall cooling effect and hence a negative value of $\Delta T/T_0$ in particular for large intruders. The cooling effect of large intruders as observed in our dense shear flow system is to some extend in contradiction to the work of Trujillo & Hermann [13] in which the intruder behaves like a heating source in a granular gas. Yet, the system studied in [13] is very different to our system, as it considers a dilute granular gas system under strong vibrations. In the system of Trujillo & Hermann [13] binary collision dominate (with a long free path length), whereas in our system multiple and enduring contacts prevail.

### C. The Continuum Limit

From Figure 9 we observe that for $d_I/d_p > 4$, $\Delta P/P_0 \to 0$. This asymptotic behaviour of $\Delta P/P_0$ is an indication that for $d_I/d_p \gg 1$ the system approaches a continuum limit, i.e. the lift force acting on the intruder approaches the value given by an Archimedean type description of the buoyancy force, i.e. $F_b = \phi \rho_p g_y V_I$, or $F_b/\rho_p g_y V_I \to \phi$ (Figure 7). This trend has also been reported by Jing et al. [20] and van der Vaart et al.

[15]. In the continuum limit, i.e. $d_I/d_p \gg 1$ the intruder behaves as if being immersed in a fluid with density $\rho_p\phi$. Jing et al. [20] argued that for $d_I/d_p \gg 1$ a large number of bed particles are surrounding and hence in contact with the intruder, leading to a high number of particle collisions and in turn a uniform stress transmission (similar to a continuum fluid). Therefore, for large values of $d_I/d_p$, $\Delta P = P_1 - P_0$ approaches zero (as confirmed in Figure 9).

To explore in more depth, the change in $P$ when an intruder in the size range $1.5 < d_I/d_p < 4$ is introduced into the system, we calculated the pressure at the location of the intruder over 100 s and plot its distribution as a function of $d_I/d_p$ in Figure 11. For $d_I/d_p = 1.5$, the modal value of the pressure is 800 Pa (mean value 987 Pa) is significantly smaller than the hydrostatic pressure of 1153 Pa. The difference between the mean value of the pressure distribution and the hydrostatic pressure is reflected in the positive pressure contribution to the upward-directed lift force acting on intruders with $d_I/d_p < 4$. In addition, the large deviation between the modal and hydrostatic pressure is likely the reason for the large fluctuations in the DEM-determined values of the lift force for smaller sizes ratios of $d_I/d_p$ (Figure 7). Such fluctuations in DEM-determined lift forces have been reported previously [15,17,20]. For $d_I/d_p = 4$, the pressure distribution becomes more symmetric with the modal ($P_{modal} = 1210$ Pa), mean ($P_{mean} = 1230$ Pa) and hydrostatic ($P_0 = 1153$ Pa) pressures being very close to each other (Figure 12(d), explaining both the small lift force and the small fluctuations in the DEM-determined lift force for $d_I/d_p \geq 4$ (Figures 7 and 9). The smaller fluctuations in the DEM-determined lift force for larger values of $d_I/d_p$ are in agreement with previous works [15,17,20].

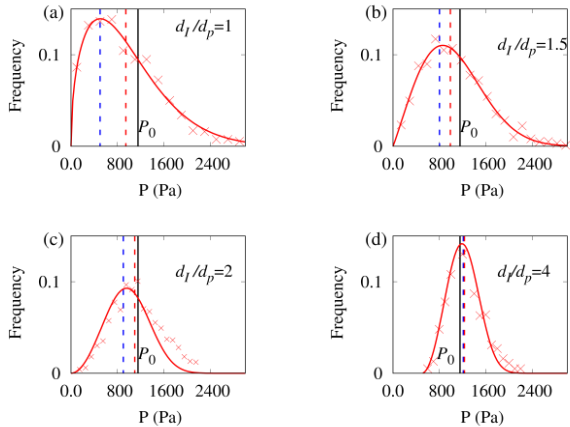

*Figure 11* The distribution of the pressure at the location of the intruder, $P(\mathbf{r}(t),t)$, as a function of the size ratio: (a) $d_I/d_p = 1$, (b) $d_I/d_p = 1.5$, (c) $d_I/d_p = 2$, and (d) $d_I/d_p = 4$. The set-up used an inclination angle of $\theta = 24°$ with $\rho_I/\rho_p = 1$. Pressure data were sampled over 100 s in steps of 0.1 s. The hydrostatic pressure $P_0 = 1153$ Pa is denoted by the black, solid vertical line, given by $P_0 = \rho_p g_y(h_c - y_0)$ with $h_c = 0.192$ m and $y_0 = 0.1$ m. The mean pressure value is denoted by the red, dashed vertical line. The modal pressure value is denoted by the blue, dashed vertical line: (a) $P_{modal} = 500$ Pa, $P_{mean} = 944$ Pa (b) $P_{modal} = 800$ Pa, $P_{mean} = 987$ Pa (c) $P_{modal} = 900$ Pa, $P_{mean} = 1097$ Pa (d) $P_{modal} = 1210$ Pa, $P_{mean} = 1230$ Pa.

## D. The Effect of Density Differences on segregation

So far we have only considered cases in which the density of the intruder and the bed particles are equal. From Eq. (31), when scaling the lift force with the intruder weight ($\rho_I g_y V_I$) the following relationship is obtained for $\rho_p \neq \rho_I$,

$$\frac{F_{\text{lift}}}{\rho_I g_y V_I} = \left(2 - e^{-\frac{\Delta T}{T_0} + \frac{\Delta P}{P_0}}\right)\frac{\rho_p}{\rho_I}. \qquad (33)$$

When $F_{\text{lift}}/(\rho_I V_I g_y) < 1$ the intruder sinks and for $>1$ the intruder rises. Similar to the model of Jing et al. [20], also Eq. (33) shows a decoupling of the effects of the intruder size and density ratios. Figure 12 plots $F_{\text{lift}}/(m_I g_y)$ as a function of $\rho_p/\rho_I$ for $d_I/d_p = 2$ and 4. The linear trend predicted by the lift force model, Eq. (33), agrees very well with the DEM data and with experimental observations that show that light intruders migrate upward while heavier intruders sink [19,41].

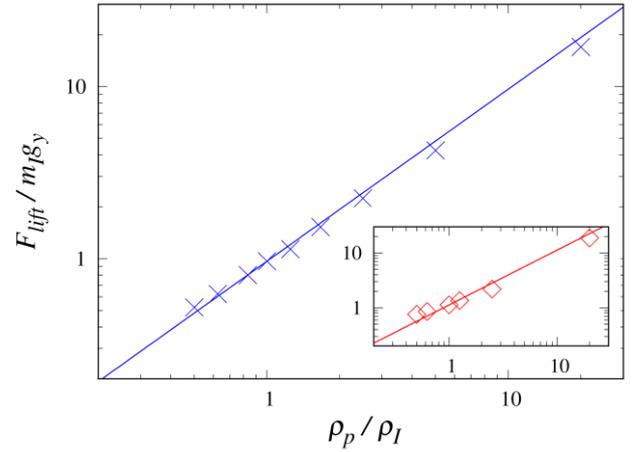

*Figure 12.* The normalized lift force as a function of the density ratio $\rho_p/\rho_I$ for different intruder sizes (×) $d_I/d_p = 4$. (◇) $d_I/d_p = 2$. The blue and red straight lines are given by Eq. (33).

From Eq. (33) it is also possible to extract the neutral buoyancy limit, which is given by $\left(2 - e^{-\frac{\Delta T}{T_0} + \frac{\Delta P}{P_0}}\right)\frac{\rho_p}{\rho_I} = 1$ and plotted in Figure 13.

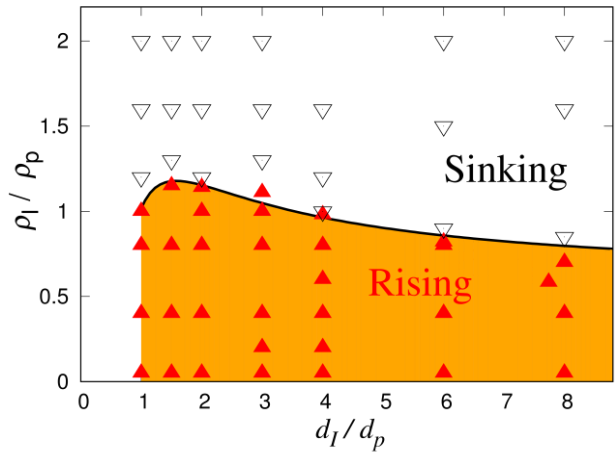

*Figure 13.* Neutral buoyancy limit of a single intruder in a dense, granular shear flow as function of $d_I/d_p$ and $\rho_I/\rho_p$. The black solid curve denotes the neutral buoyancy limit given by $\left(2 - e^{-\frac{\Delta T}{T_0} + \frac{\Delta P}{P_0}}\right) \frac{\rho_p}{\rho_I} = 1$. DEM data using an inclination angle of $\theta = 25°$. $\triangledown$ intruder sinks, ▲ intruder rises.

## 4. Conclusions

In this work, we propose a new lift force model for intruders in dense, granular shear flows by extending the work of Trujillo & Hermann's [13]. The lift force is interpreted as a buoyancy force whereby the density difference arises both from granular temperature and granular pressure contributions. We observe that the presence of an intruder leads to a cooling effect and a local flattening of the shear velocity profile (lower shear rate). For large intruders, i.e. $d_I/d_p > 4$, the local pressure disturbance (and hence contribution to the lift force) is very small as the system approaches a continuum limit, in which the pressure acting on the intruder equals to the hydrostatic pressure of the system. On the other hand, for $1 < d_I/d_p < 4$ the local granular pressure at the location of the intruder is lower than the hydrostatic pressure leading in turn to a positive lift force. The cooling effect due to the presence of an intruder increases with intruder size, leading ultimately to the sinking of large intruders. The modified model predicts DEM-determined lift forces very well and allows the determination of a neutral buoyancy limit.

## Acknowledgment
We are thankful to the Swiss National Science Foundation (Grant No. 200020_182692), Fondation Claude et Giuliana and the China Scholarship Council (M.L.) for partial financial support of this work. This publication was created as part of NCCR Catalysis, a National Centre of Competence in Research funded by the Swiss National Science Foundation. We acknowledge J.P. Metzger and N.A. Conzelmann for useful comments while preparing the manuscript.